\documentclass[aip,reprint,amsmath,amssymb,nofootinbib,superscriptaddress]{revtex4-2}
\usepackage{graphicx}
\usepackage{dcolumn}
\usepackage{bm}
\usepackage{amsmath,amssymb}
\usepackage{epstopdf}
\usepackage{color}
\usepackage{verbatim}
\usepackage{caption}
\usepackage{subcaption}
\usepackage{float}
\usepackage{ulem}

\begin{document}
\title{Spatially Resolved Reconstruction of Ising Couplings in Tunable Colloidal Artificial Spin Lattices}
\author{Qingyu Qu$^{\dag}$\footnotemark[1]}
\affiliation{Department of Physics and Center for Complex Flow and Soft Matter Research, Southern University of Science and Technology, Shenzhen, Guangdong 518055, China}
\author{Yongming Zhang$^{\dag}$\footnotemark[1]}
\affiliation{Department of Physics and Center for Complex Flow and Soft Matter Research, Southern University of Science and Technology, Shenzhen, Guangdong 518055, China}
\author{Xiaoguang Ma}
\email{maxg@sustech.edu.cn}
\affiliation{Department of Physics and Center for Complex Flow and Soft Matter Research, Southern University of Science and Technology, Shenzhen, Guangdong 518055, China}
\renewcommand\thefootnote{}
\footnotetext[1]{$^\dag$ These authors contributed equally.}
\date{\today}
\begin{abstract}
Buckled colloidal monolayers constitute a versatile soft-matter platform for engineering artificial spin lattices, with each particle serving as a single Ising spin. While the average Ising coupling energy has been approximately derived for perfect particle lattices, extracting the complete set of coupling parameters from real samples remains inaccessible. 
Here, we apply an inverse method to reconstruct all nearest-neighbor effective Ising coupling energies from measured colloidal spin configurations. We design multiple experimental protocols to control the thermodynamic state of the colloidal system, from isotropic compression and shear deformation to modulation of interparticle attraction, each giving rise to distinct spin configurations. Using spin configuration data, we reconstruct all nearest-neighbor effective Ising coupling energies using maximum likelihood estimation. To assess inference reliability without ground-truth model parameters, we propose to use the convergence of the standard deviation of the estimated couplings as a practical, ground-truth-free criterion, and validate its reliability using simulation data with known parameters. The extracted spatially resolved couplings reveal how each control protocol influences the sign, magnitude, statistical distribution, and spatial arrangement of the microscopic coupling parameters: isotropic compression strengthens antiferromagnetic couplings and enhances quenched disorder; shear deformation generates direction-dependent anisotropic couplings; and increased interparticle attraction drives a crossover from antiferromagnetic to paramagnetic and then to ferromagnetic couplings. This work establishes a practical inference framework for estimating effective model parameters of colloidal artificial spin lattices.
\end{abstract}
\maketitle

\section{Introduction}
\label{sec:introduction}
Synthetic artificial spin lattices (ASLs) comprise engineered assemblies of interacting units whose energies can be mapped to a coupled spin model \cite {2006.wang,2008.han,2013.geim,2016.tierno,2025.wang,jvzh-lwbh}. ASLs can be designed and synthesized using patterned magnetic nanoislands, van der Waals heterostructures, and colloidal bead arrays; these synthetic systems can mimic the physical behavior of a variety of spin models. Thanks to their mesoscopic dimensions and design flexibility, ASLs offer many advantages over electronic and atomic spin systems as they permit direct real-space visualization of individual spins, support customizable geometry and topology, and enable site-resolved control over individual spins and inter-unit interactions. These convenient features make ASLs ideal model systems for investigating exotic physical phenomena, from spin frustration \cite{2006.wang,2008.han,2009.shokef,2014.gilbert,2016.tierno,2016.ortiz,2017.zhou,jvzh-lwbh}, topological excitations \cite{2010.ladak,2011.mengotti,2026.baillou}, skyrmions \cite{2019.han,2020.ding,2018.tong,2020.wu}, to collective spin-wave dynamics \cite{2004.gubbiotti,2005.bayer,2018.mamica,2018.monton,2012.tierno}. For instance, patterned magnetic nanoislands are used to construct artificial spin ice, which replicates the behavior of geometrically frustrated atomic magnetic spins \cite{1997.harris}. Beyond fundamental interests, ASLs also hold promise for next-generation computing; ASLs have proven efficient for solving complex combinatorial optimization problems through Ising-style analog computation \cite{2018.jensen,2019.arava,2021.psaroudaki,2022.mohseni,2023.hu}.

Among various realizations, colloidal ASLs are soft-matter systems whose energies are close to the thermal energy, and are far lower than those of magnetic systems. This low energy scale enables colloids to sensitively respond to thermal excitations and external perturbations, which is a desired property in exploring equilibrium phase behavior and designing tunable matter \cite{1989.tang,1998.wei,2001.bechinger,2008.mikhael,2014.alert,2016.kakoty,jvzh-lwbh}. For example, paramagnetic colloidal beads feature tunable magnetic moments. Studies leveraging these adjustable interactions have revealed distinct phase transition behavior at the mesoscopic scale \cite{2014.alert}.

Buckled colloidal monolayers are a unique class of colloidal ASLs \cite{2008.han,2009.shokef,2017.leoni,2023.hill,jvzh-lwbh}. Governed purely by repulsive hard-sphere interactions, the closely-packed, quasi-two-dimensional particle assemblies exhibit effective Ising-like spins and couplings that can be experimentally tuned \cite{2017.leoni,2019.ma,2023.hill}. For instance, an increase in interparticle attraction can drive the ASLs to transition from an antiferromagnetic to a paramagnetic and then to a ferromagnetic phase \cite{2023.hill}. Recently, optical tweezers have been integrated into the buckled particle system. The programmable positions and powers of optical tweezers permit reconfiguration of lattice geometry and in-situ mechanical deformation, opening up new strategies for manipulating these colloidal ASLs \cite{jvzh-lwbh}.

Despite experimental advances, quantitative characterization of the effective spin model to which these ASLs are mapped remains incomplete. In colloidal ASLs, for example, while the Ising couplings of the buckled colloids have been derived from particle parameters, these calculations are approximate and their predictions sensitively depend on the real parameters (e.g., diameter and surface charge) of the experiment, which are impossible to accurately measure. As a result, there is no direct comparison between theoretical calculations and experiments \cite{2008.han,2023.hill}. Another challenge is to extract not just the average coupling but the complete set of spatially resolved coupling energies across the sample. The latter task is critical for analyzing ASLs with nonuniform parameters. Even for uniform ASLs, access to the entire set of parameters is required to characterize the uniformity.

Thus, inverse methods that reconstruct model parameters from data are a promising solution \cite{2004.paninski,2011.sohl,2012.aurell,2017.nguyen,2023.kloucek}. Most inverse methods have been benchmarked against datasets numerically generated from models with known parameters. Applying these inverse methods to real ASLs, however, is problematic as most experiments lack known ground-truth parameters. It is not clear how to judge whether a given experimental dataset is sufficient for reliable inverse inference.

In this work, we apply maximum likelihood estimation (MLE), a well-known inverse method, to spin configuration data measured from buckled colloidal ASLs. We first use simulated spin configurations with known parameters to establish that the standard deviation of the estimated couplings converges to its true value when the dataset size is sufficiently large. We thus propose to use this convergence as a practical, ground-truth-free criterion. Then, using MLE with this new criterion, we show that tuning various thermodynamic states of the colloidal ASLs, including isotropic compression, shear deformation, and temperature-dependent depletion attraction, allows systematic control of the effective Ising couplings. The extracted spatially resolved couplings reveal how each macroscopic stimulus alters the sign, magnitude, statistical distribution, and spatial arrangement of the microscopic spin couplings. This practical inference framework can be applied to other ASLs as well.

The remainder of the paper is organized as follows. Section \ref{sec:experiment} describes the preparation and tuning protocols of the colloidal ASLs. Section \ref{sec:mle} explains the MLE and proposes a ground-truth-free convergence criterion. Section \ref{sec:result} presents the experimental data and the inferred coupling parameters for different tuning protocols. Finally, Sec.~\ref{sec:summary} summarizes the work.

\section{Experiment}
\label{sec:experiment}
We use polystyrene microspheres with a nominal diameter $D = 1.0 \pm 0.1~\mu\text{m}$ (Thermo Fisher Scientific) to prepare the sample. We first clean the particle solution and remove impurities following the method in Ref. \cite{jvzh-lwbh}. The sample is contained in a home-made transparent sample cell consisting of two cover glasses ($22 \times 40~\text{mm}$, Thermo Fisher Scientific). The cover glasses are thoroughly cleaned before use. The sample has a wedge shape with an angle of about $0.001~\text{rad}$. Thus, the glass surfaces are close to being parallel. We inject $20~\mu\text{L}$ of diluted, purified colloidal suspension into the wedge cell and completely seal the cell using optical adhesive (Norland 65) under UV illumination. 

\begin{figure}[htbp]
	\centering
	\includegraphics[width=0.8\linewidth]{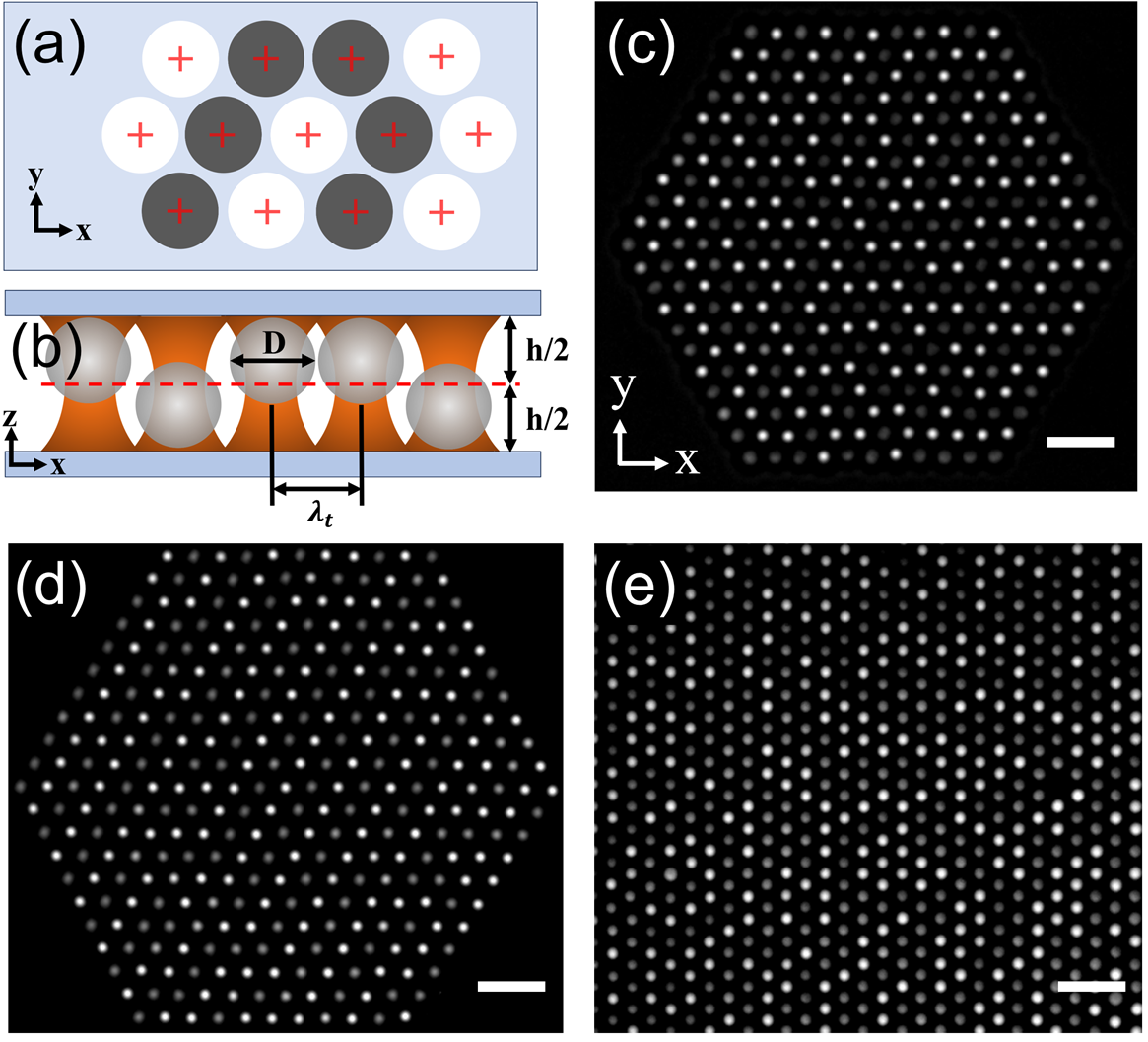}
	\caption{(a) Schematic of the sample. Gray spheres: particles. Blue rectangles: cover glass. Orange beams: optical tweezers. (b) Top view of (a). Red crosses: optical tweezers. Snapshots from three experiments: (c) Compression, (d) Attraction tuning, and (e) Shear.}
	\label{fig:experiment}
\end{figure}

We use an inverted optical microscope (Ti2U, Nikon) for observation. The microscope is equipped with a $100\times$ oil-immersion objective (N.A. $= 1.2$) and a $1920 \times 1280$ high-resolution camera. Our optical tweezers system (Tweez 305, Aeresis) is also integrated with this microscope. We record digital images at $20$ fps, with a typical acquisition time of $30$ minutes. We generate $331$ optical tweezers arranged into a triangular lattice, and put a colloidal particle in each optical trap with an effective laser power of 1 mW.

To compress the particle lattice, we reduce the lattice constant $\lambda_t$ of the optical tweezers array \cite{jvzh-lwbh}. To shear the particle lattice, we apply an affine shear transformation to the lattice of optical traps \cite{jvzh-lwbh}. See Figs.~\ref{fig:experiment}(b,c).  

For the attraction-tuning experiment, we suspend colloidal particles in $44~\text{mM}$ hexaethylene glycol monododecyl ether ($\mathrm{C_{12}E_6}$) and $2~\text{mM}$ NaCl solution. Colloidal particles exhibit short-range depletion attractions which increase with sample temperature \cite{2019.ma, 1996.dinsmore, 2016.gratale, 2023.hill}. A snapshot of a temperature-tunable sample is shown in Fig.~\ref{fig:experiment}(d).

\section{Maximum Likelihood Estimation}
\label{sec:mle}
We briefly explain maximum likelihood estimation (MLE). The energy of an Ising model in the absence of external fields is
\begin{equation}
	H = - \sum_{\langle i,j \rangle} J_{ij} s_i s_j,
	\label{eq:ising}
\end{equation}
where $s_i, s_j \in \{+1, -1\}$ denote Ising spins, $J_{ij}$ denotes couplings between the $i$-th and $j$-th spin. The summation is over all nearest-neighbor spin pairs.

In equilibrium, the probability of a spin configuration $\mathbf{s}\equiv\{s_1,s_2,\dots,s_N\}$ follows the Boltzmann distribution:
\begin{equation}
	P(\mathbf{s}\mid \{J_{ij}\}) = \frac{1}{Z}\exp\left(-\beta H \right),
	\label{eq:boltzmann}
\end{equation}
where $\beta\equiv (k_B T)^{-1}$ is the inverse thermal energy, and $Z\equiv \sum_{\mathbf{s}} \exp\left(-\beta H \right)$ is the canonical partition function. Equation~\eqref{eq:boltzmann} defines the mapping from model parameters $\mathbf{J}\equiv \{J_{ij}\}$ to the statistical distribution of spin configurations $P(\mathbf{s})$; this is known as the forward Ising problem.

Given a dataset, $S = \{\mathbf{s}^{(1)}, \mathbf{s}^{(2)}, \ldots, \mathbf{s}^{(M)}\}$, that consists of $M$ spin configurations, MLE infers the optimal parameters $\hat{\mathbf{J}}$ that best reproduce the statistical properties of the data by maximizing the function:
\begin{equation}
	\hat{\mathbf{J}} = \arg\max_{\mathbf{J}} L(\mathbf{J}),
	\label{eq:mle_def}
\end{equation}
where $L(\mathbf{J}) = \log P(S \mid \mathbf{J})$ is the log-likelihood of observing the data $S$ under the couplings $\mathbf{J}$.

For a finite dataset $S$, the maximizer of $L(\mathbf{J})$ in Eq.~\eqref{eq:mle_def} is computed via a gradient ascent algorithm, starting from an initial guess of $\mathbf{J}$ (e.g., all zeros or small random values). Then, we iteratively update each coupling parameter along the direction of the log-likelihood gradient:
\begin{equation}
	\frac{\partial L(\mathbf{J})}{\partial J_{ij}}
	= \beta \bigl(\langle s_i s_j \rangle_S
	- \langle s_i s_j \rangle_{\text{ens}}(\mathbf{J})\bigr),
	\label{eq:gradient}
\end{equation}
where $\langle \cdot \rangle_S$ denotes the empirical average over the input dataset $S$, and $\langle \cdot \rangle_{\text{ens}}$ denotes the thermal ensemble average computed from Eq.~\eqref{eq:boltzmann} with the current estimate of $\mathbf{J}$. In practice, the ensemble average is approximated by an empirical average over configurations generated by equilibrium Monte Carlo (MC) simulation using the current $\mathbf{J}$. By repeating the gradient update step with a suitable learning rate, the coupling parameters $\mathbf{J}$ converge to the maximum-likelihood estimate $\hat{\mathbf{J}}$.

To evaluate the accuracy of MLE, simulation data has been used for benchmarking \cite{2012.aurell,2017.nguyen}, where estimated couplings $\hat{\mathbf{J}}$ are compared against ground-truth parameters used to generate the data. Previous work has identified dataset size $M$ and the characteristic energy scale of the ground-truth couplings $\langle J_{ij}\rangle$ as the two major factors that influence the estimation accuracy: smaller datasets and stronger couplings both lead to larger reconstruction errors.

For experimental datasets, however, ground-truth parameters are in general inaccessible. A ground-truth-free metric is thus necessary for evaluating the performance of MLE. To this end, we first reproduce the benchmark test with simulations tailored to our experimental system; we then compare the ground-truth-based metric with other metrics that do not require ground-truth parameters. 

We perform canonical MC simulations of Eq.~\eqref{eq:ising} on a two-dimensional triangular lattice with $N = 331$ spins, matching our experimental setup. For each simulation, we generate coupling parameters $J_{ij}$ sampled from a normal distribution with different means $\langle J \rangle$ and standard deviation $\delta J$. Each dataset consists of $M$ equilibrium spin configurations extracted from the simulation. We then apply MLE to each dataset to reconstruct $\hat{\mathbf{J}}$.

Following Ref.~\cite{2017.nguyen}, we quantify the normalized reconstruction error by
\begin{equation}
	\gamma_J = \sqrt{\frac{\sum_{\langle i,j \rangle} (\hat{J}_{ij} - J_{ij})^2}{\sum_{\langle i,j \rangle} J_{ij}^2}}.
	\label{eq:rec_err}
\end{equation}
$\gamma_J$ measures the relative difference between reconstructed couplings $\hat{\mathbf{J}}$ and the ground-truth ones $\mathbf{J}$.

\begin{figure}[htbp]
	\centering
	\includegraphics[width = 0.7\linewidth]{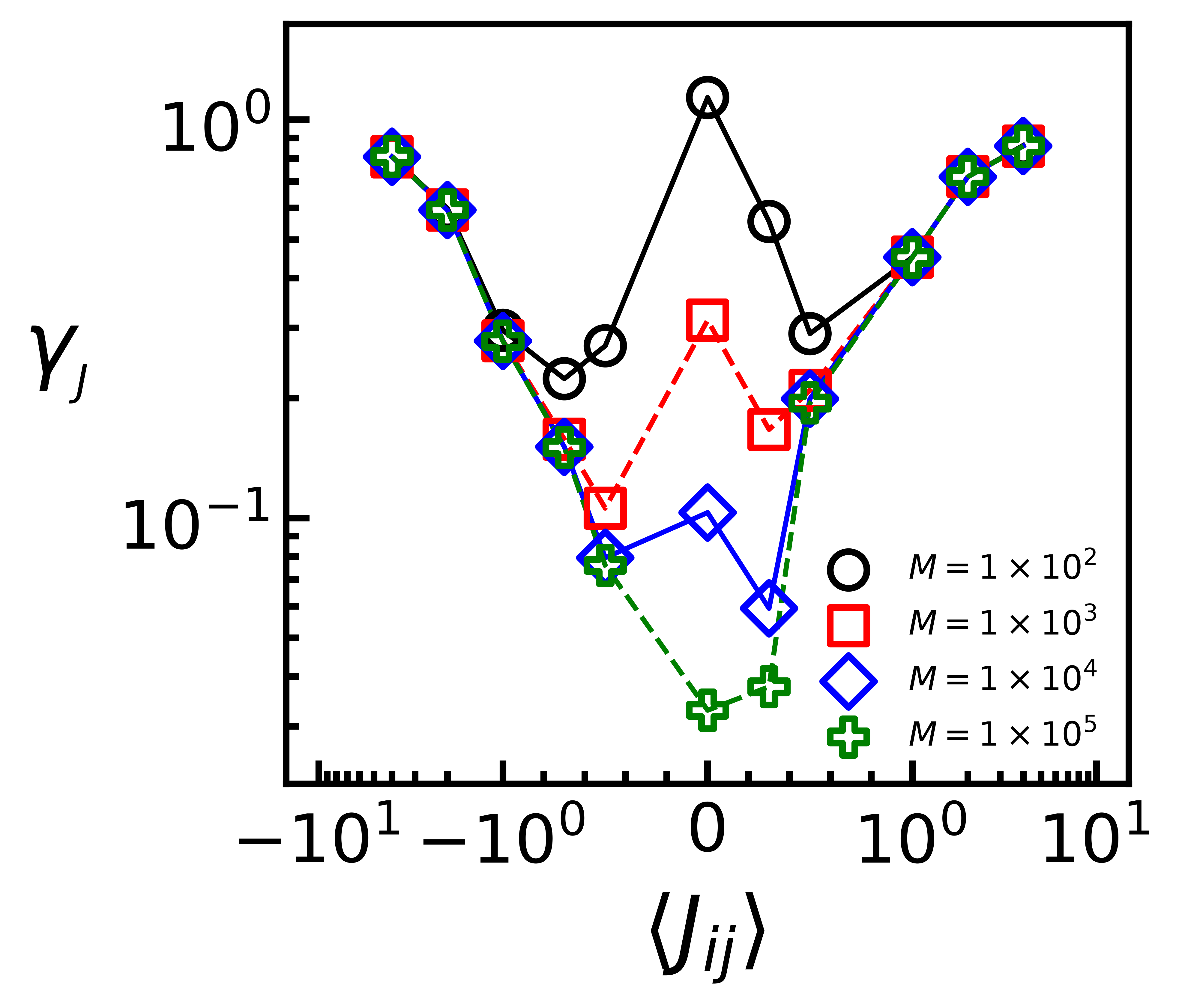}
	\caption{Normalized estimation error $\gamma_{J}$ as a function of mean coupling strength $\langle J_{ij} \rangle$ for datasets of different sizes $M$.}
	\label{fig:accuracy}
\end{figure}

As expected, the error $\gamma_J$ decreases monotonically with increasing dataset size $M$, as shown in Fig.~\ref{fig:accuracy}. However, the rate at which $\gamma_J$ decreases with $M$ depends on the magnitude of the coupling energies characterized by, e.g., $\langle J\rangle$. When $\langle J\rangle$ is close to zero, $\gamma_J$ rapidly drops below $0.1$ for $M \simeq 1 \times 10^4$. By comparison, for $|\langle J\rangle|\geq 1$, $\gamma_J$ remains large (e.g., $\gamma_J>0.2$) and decays much more slowly. This comparison indicates the accuracy behavior of MLE: for reliable reconstruction with $\gamma_J<0.1$, a necessary condition is that $-1< \langle J\rangle <1$, and the dataset size should be $M > 1 \times 10^4$.

\begin{figure}[htbp]
	\centering
	\includegraphics[width=0.7\linewidth]{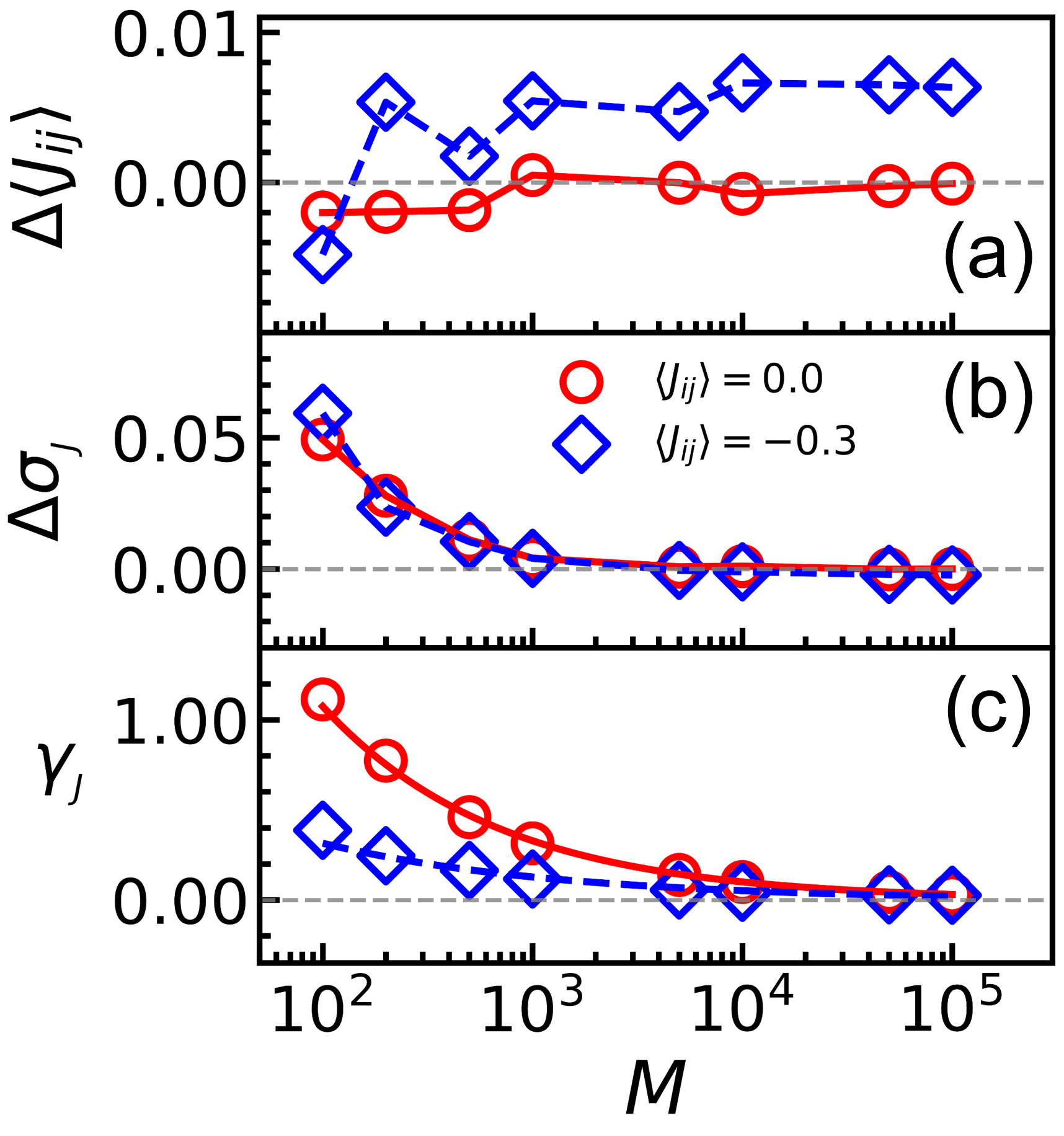}
	\caption{
		(a) Normalized estimation error $\gamma_J$ as a function of dataset size $M$.
		(b) Deviation of the mean estimated coupling from the ground-truth mean, $\Delta\langle J\rangle$, as a function of dataset size $M$.
		(c) Deviation of the standard deviation of estimated couplings from the ground-truth value, $\Delta\sigma_J$, as a function of dataset size $M$.
		Blue diamonds: $\langle J\rangle=-0.30$. Red circles: $\langle J\rangle=0.00$.
	}
	\label{fig:simulation_mean_std}
\end{figure}

For experimental datasets, $\gamma_J$ cannot be computed since the ground-truth couplings are unknown. We thus propose alternative metrics that derive solely from accessible data. To this end, we evaluate two candidate metrics: the mean $\langle \hat{J}\rangle$ and the standard deviation $\sigma_{\hat{J}}$ of the estimated couplings. We examine their convergence behavior as a function of dataset size for two different ground-truth coupling sets with $\langle J \rangle=0.00$ and $-0.30$, respectively. We compare the accuracy behavior of these alternative metrics with $\gamma_J$ and determine their performance.

For the case $\langle J \rangle=0.00$, the difference, $\Delta \langle J\rangle \equiv \langle \hat{J}\rangle - \langle J\rangle$, between the means of estimated and ground-truth couplings remains near zero across all tested dataset sizes [Fig.~\ref{fig:simulation_mean_std}(a)]. For the stronger coupling set with $\langle J \rangle=-0.30$, $\Delta \langle J\rangle$ asymptotically plateaus at a small but nonzero value (approximately $0.006$). The onset of this plateau appears to coincide with the point where $\gamma_J\simeq 0.1$. However, no such coincidence is observed for the case $\langle J \rangle=0.00$, suggesting that this metric is sensitive to the magnitude of coupling strength.

By comparison, the difference, $\Delta \sigma_J \equiv \sigma_{\hat{J}} - \sigma_J$, between the standard deviations of estimated and ground-truth couplings exhibits a more consistent convergence behavior for both coupling sets. As shown in Fig.~\ref{fig:simulation_mean_std}(b), $\sigma_{\hat{J}}$ is consistently larger than $\sigma_J$; the difference $\Delta \sigma_J$ monotonically decreases with $M$ and approaches zero at large $M$ for both values of $\langle J \rangle$. 

Note that we also test datasets generated from other $\langle J \rangle$ values within $[-1,1]$ and observe a similar $\Delta \sigma_J$ plateau for large $M$. Therefore, $\sigma_{\hat{J}}$ appears to be a more robust metric than $\langle \hat{J}\rangle$, as it readily converges to an asymptotic $\sigma_J$. This convergence behavior closely mirrors that of $\gamma_J$ as well. For sufficiently large datasets, both cases with $\langle J \rangle=0.0$ and $-0.3$ fall within the reliable estimation regime: $\gamma_J$ drops below 0.1 for $M>1 \times 10^4$ in both cases [Fig.~\ref{fig:simulation_mean_std}(c)]. 

The above analysis suggests $\sigma_{\hat{J}}$ can be used as a practical ground-truth-free criterion for model reconstruction when $\gamma_J$ is inaccessible. For reliable inference, $\sigma_{\hat{J}}$ should plateau within the size of the data; otherwise inferred parameters are inaccurate. For example, if $\sigma_{\hat{J}}$ is observed to decrease throughout the entire data size range, the estimation may not have converged, indicating the dataset is insufficient. Furthermore, for the best accuracy, application of MLE should be limited to models whose coupling energies are no larger than the thermal energy (e.g., $|J_{ij}|/k_BT<1$). If estimated couplings are found to satisfy $|\hat{J}_{ij}|/k_BT>1$, the reconstruction error is expected to be large and MLE may not be at all reliable. Thus, both $\sigma_{\hat{J}}$ and $J_{ij}$ values should be considered in the assessment of reconstruction.

\section{Results and discussion}
\label{sec:result}
\subsection{Estimation accuracy for experimental data}
We focus on experimental datasets in this section. We first present the general convergence behavior of $\sigma_{\hat{J}}$ obtained from our colloidal ASLs. We then incorporate this criterion into MLE, and discuss reconstructed coupling parameters across the three experiments.    

\begin{figure}[htbp]
	\centering
	\includegraphics[width=0.7\linewidth]{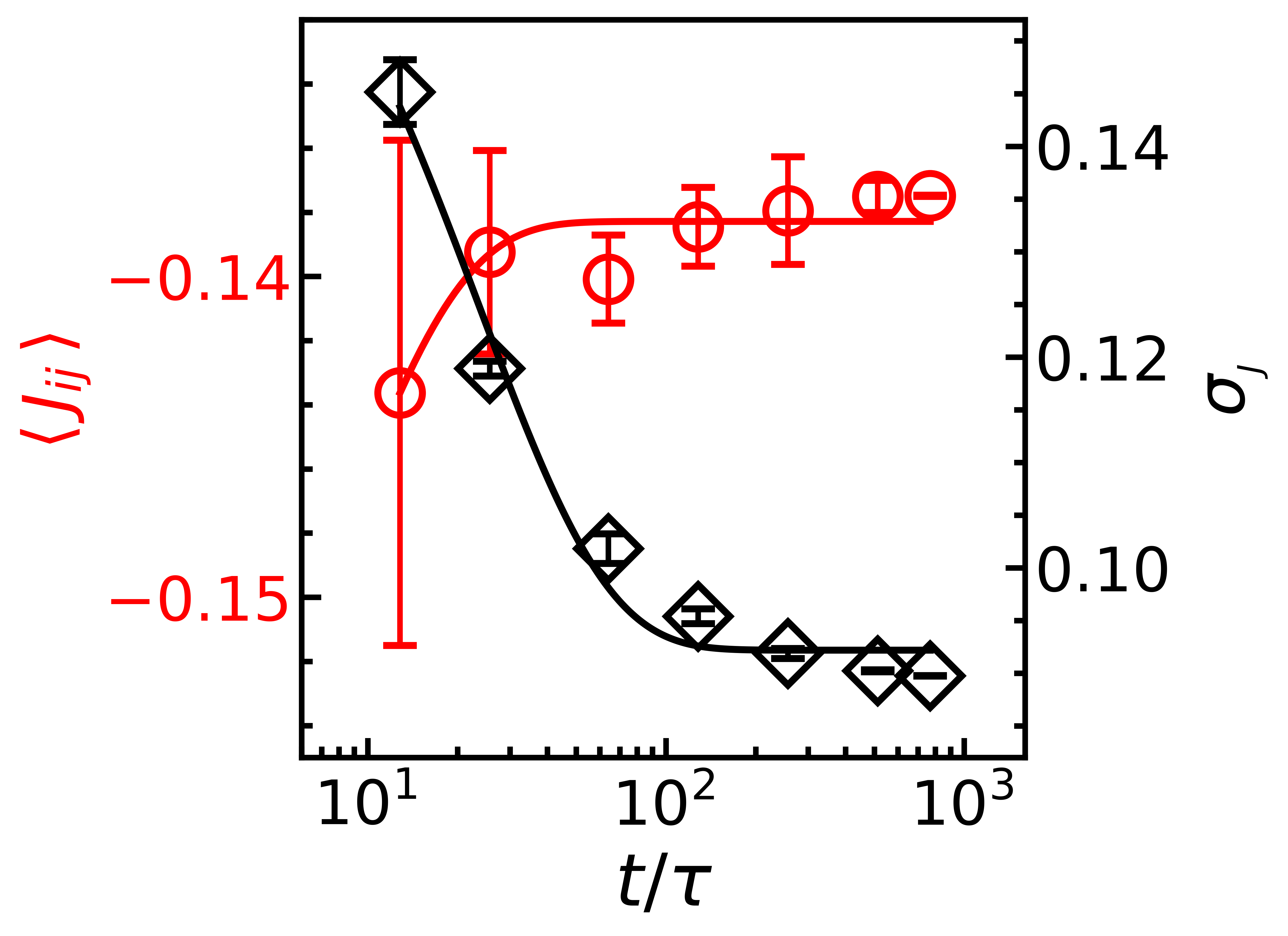}
	\caption{Histograms of estimated couplings $\rho(\hat{J}_{ij})$ for different dataset durations: $t/\tau \simeq 13$, $132$, $395$, and $526$.}
	\label{fig:num_frame_sweep}
\end{figure}

\begin{figure}[htbp]
	\centering
	\includegraphics[width=0.7\linewidth]{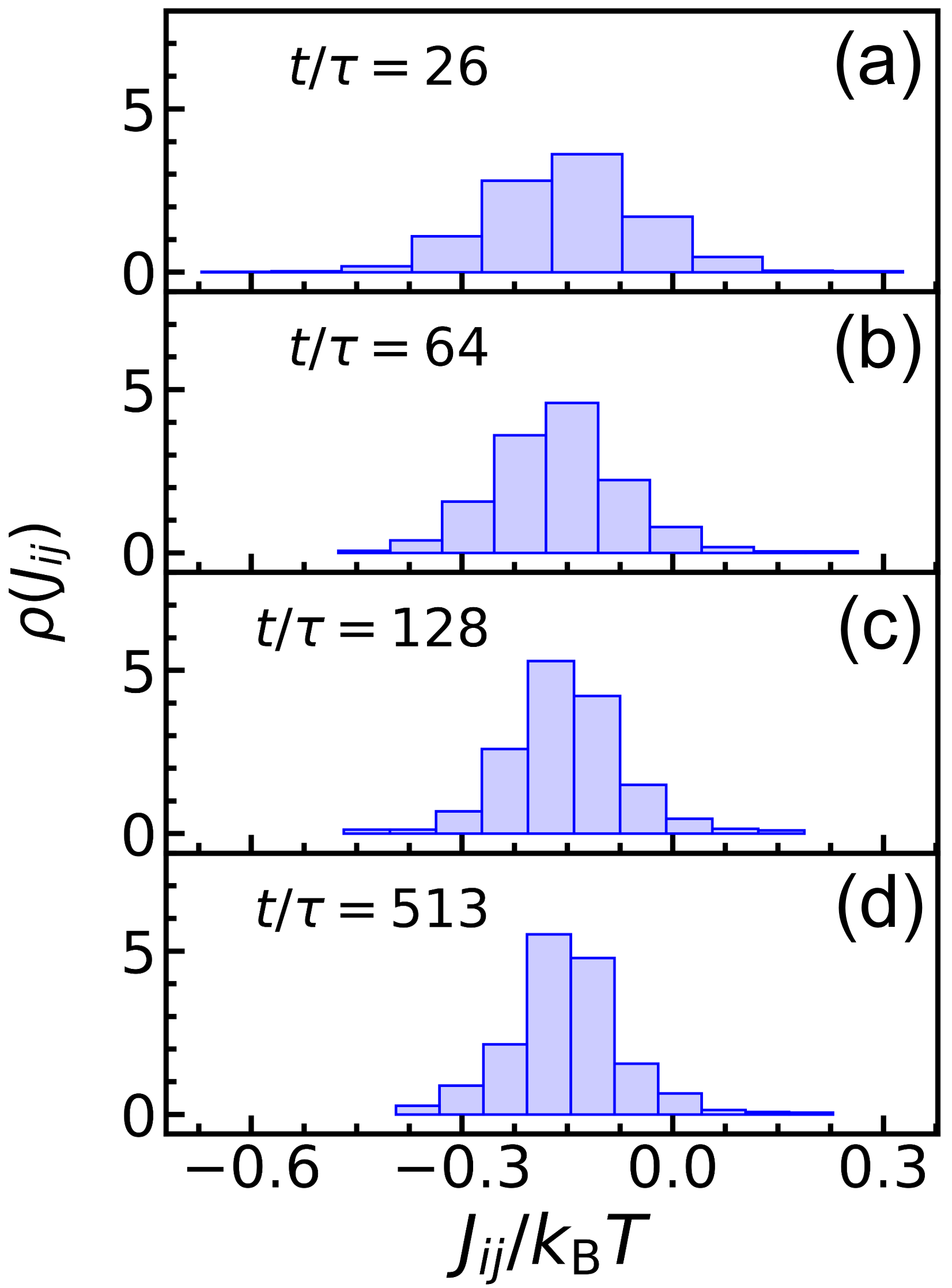}
	\caption{Mean $\langle J_{ij}\rangle$ and standard deviation $\sigma_J$ of estimated $\{J_{ij}\}$ vs dataset duration $t$. Black curves are exponential fits, $-0.09 - 0.05 \, e^{-x / 90.47}$ and $0.06 + 0.36 \, e^{-x / 31.67}$, for $\langle J_{ij}\rangle/k_BT$ and $\sigma_J/k_BT$, respectively.}
	\label{fig:num_frame_sweep_mean_std}
\end{figure}

We choose a dataset from the compressed sample with $\lambda/D\simeq 1.11$. The relaxation time measured from the spin autocorrelation function is $\tau=0.38$ s \cite{jvzh-lwbh}. From the same dataset, we apply MLE using sub-datasets with durations from $t/\tau\simeq 13$ to $526$. Figure~\ref{fig:num_frame_sweep_mean_std} plots the probability distribution density of estimated couplings $\rho(J_{ij})$ for different $t$. For the shortest $t/\tau\simeq 13$, $\rho(J_{ij})$ exhibits a broad distribution with a standard deviation $\sigma_J/k_BT\simeq 0.2$ [Fig.~\ref{fig:num_frame_sweep_mean_std}(a)]. When $t/\tau$ increases to about $132$, the width of $\rho(J_{ij})$ becomes clearly narrower [Fig.~\ref{fig:num_frame_sweep_mean_std}(b)]. This change due to larger dataset size thus indicates both datasets are insufficient. For the largest two data sets with $t/\tau\simeq 395$ and $526$, however, we observe similar $\rho(J_{ij})$, suggesting these datasets may be sufficient.

Figure~\ref{fig:num_frame_sweep_mean_std} plots the mean $\langle J_{ij}\rangle$ and standard deviation $\sigma_J$ of estimated $\{J_{ij}\}$ for various dataset durations $t$. Both $\langle J_{ij}\rangle$ and $\sigma_J$ approach their respective asymptotic values when $t/\tau>200$. We fit $\langle J_{ij}\rangle$ and $\sigma_J$ vs $t$ using the exponential function, $x = a + be^{-t/s}$ ($x=\langle J_{ij}\rangle$ or $\sigma_J$), yielding $\{a=-0.09, b=-0.05, s=90\}$ for $\langle J_{ij}\rangle$ and $\{a=0.06, b=0.36, s=90\}$ for $\sigma_J$, respectively. Based on the comparison between $\gamma_J$ and $\sigma_J$ using simulation data, we expect the reconstruction error for the experimental data to become negligibly small (e.g., $\gamma_J<0.1$) when $\sigma_J$ reaches the asymptotic value $a=0.06$. Therefore, $\sigma_J$ vs $t$ serves as a practical metric to determine whether the duration of experimental data is sufficient for reliable estimation of the couplings; when $\sigma_J$ no longer decreases with $t$, MLE generates an optimal reconstruction result.

In the following, we show reconstruction results for the three experiments. We ascertain that all results are obtained provided that $\sigma_J$ plateaus within the dataset duration. 

\subsection{Estimated couplings from compressed samples}
We first show reconstructed coupling parameters from the compressed samples with $\lambda/D = 1.11$, $1.10$, and $1.09$. Because the relaxation time is longer when the sample is more compressed, we use correspondingly longer dataset durations of $300$, $300$, and $600$ s, respectively. $\sigma_J$ is verified to plateau within these durations.

\begin{figure}[htbp]
	\centering
	\includegraphics[width=1\linewidth]{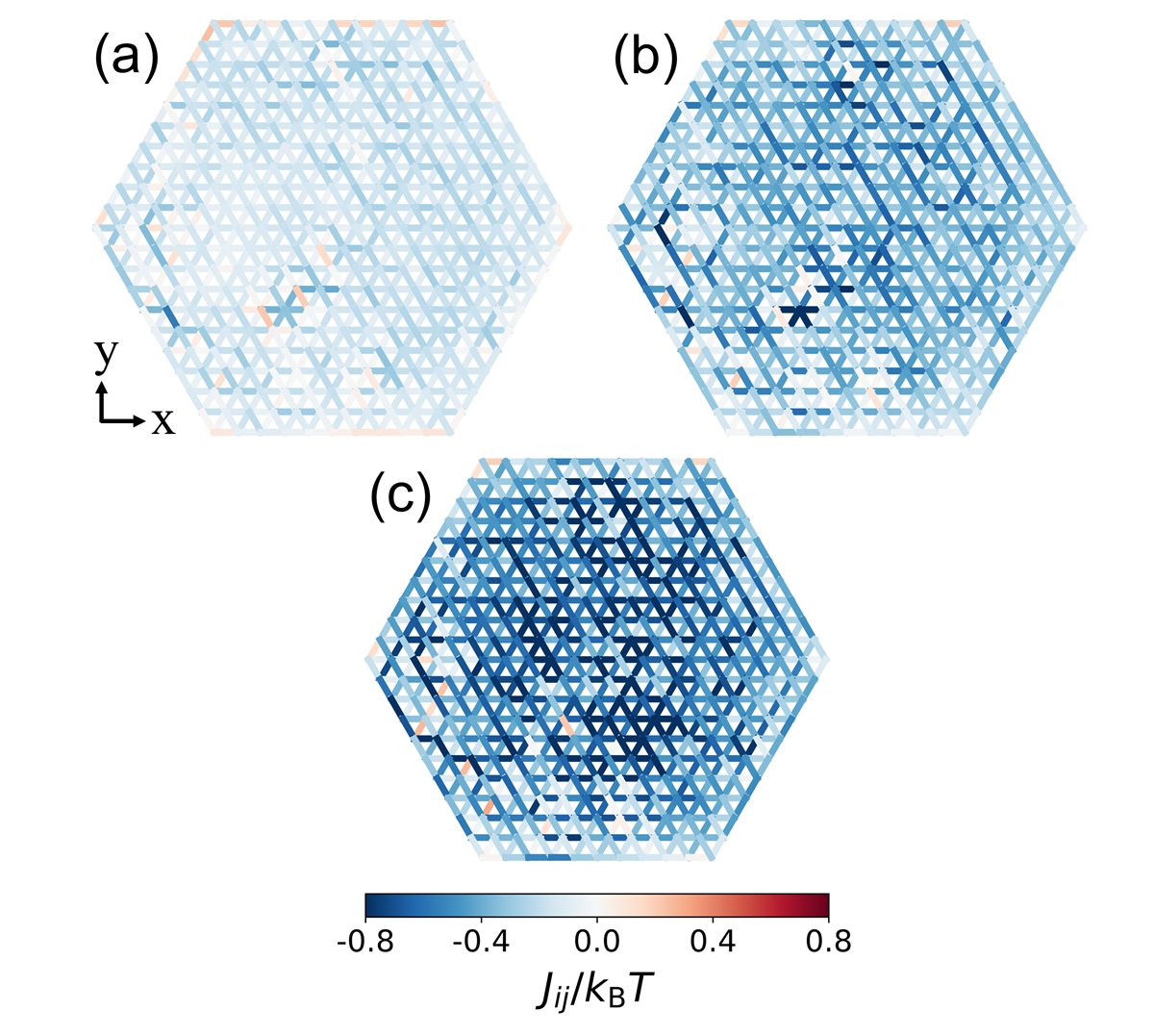}
	\caption{Spatial arrangement of estimated $\{J_{ij}\}$ from compressed samples for (a) $\lambda/D = 1.11$, (b) $\lambda/D = 1.10$, and (c) $\lambda/D = 1.09$. Bars: individual $J_{ij}$'s; Bar color: coupling strength ($J_{ij}/k_BT$).}
	\label{fig:compression}
\end{figure}

\begin{figure}[htbp]
	\centering
	\includegraphics[width=1.0\linewidth]{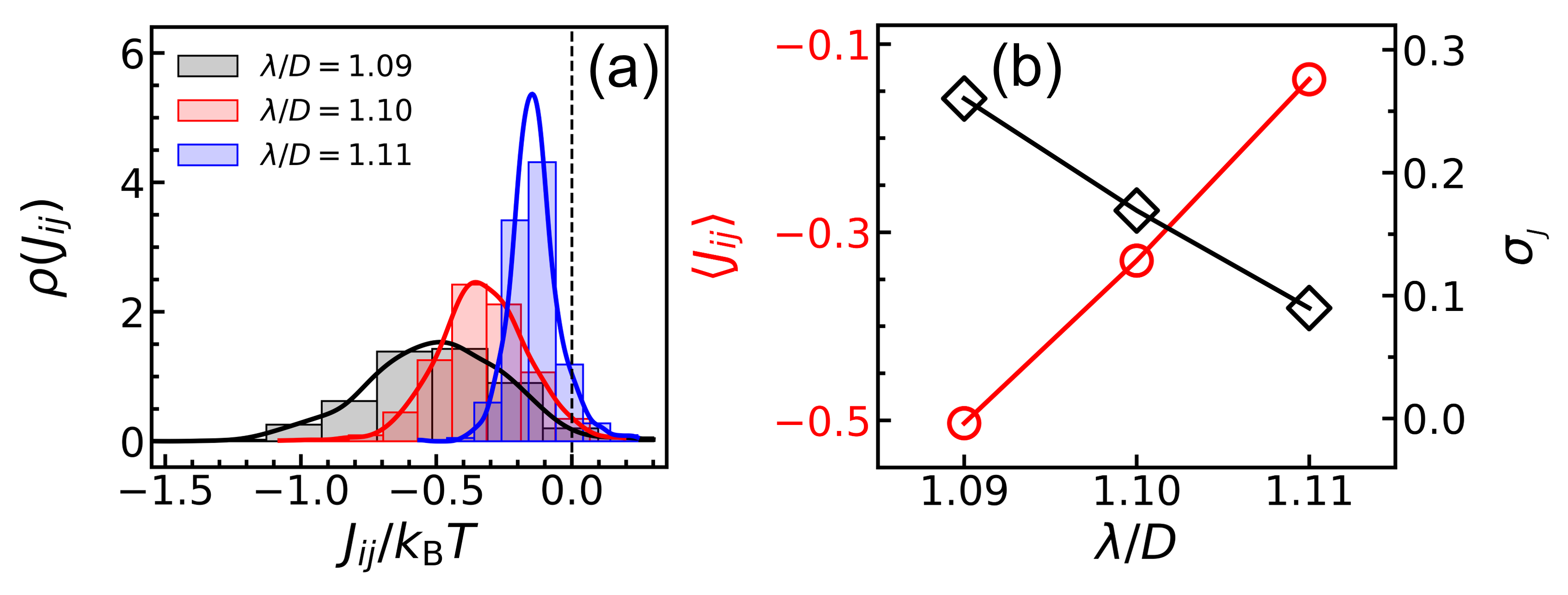}
	\caption{(a) Histograms of estimated couplings $\rho(J_{ij})$ for compressed samples. (b) Mean $\langle J_{ij}\rangle$ and standard deviation $\sigma_J$ of $\rho(J_{ij})$. Solid curves are best Gaussian fits.}
	\label{fig:compression_histogram}
\end{figure}

Figure~\ref{fig:compression} plots the spatial arrangement of estimated $\{J_{ij}\}$ for the three samples with $\lambda/D = 1.11$, $1.10$, and $1.09$ in the compression experiment. $\{J_{ij}\}$ are represented by short bars connecting nearest-neighbor sites, with the color indicating the sign and magnitude of the coupling energy (red: positive or ferromagnetic coupling, blue: negative or antiferromagnetic coupling, white: no coupling). For the least compressed sample ($\lambda/D = 1.11$), although a few couplings exhibit positive values (red), the overall coupling energies are negative and close to zero (light blue) [Fig.~\ref{fig:compression}(a)]. For samples with smaller lattice constants ($\lambda/D = 1.10$ and $1.09$), $\{J_{ij}\}$ become more negative. This increase in antiferromagnetic (negative) couplings due to stronger compression is consistent with previous experiments, where the mean coupling has been derived from the free volume of particles in buckled states  \cite{2008.han,2009.shokef,2023.hill}. Nevertheless, using the inverse method we are able to reconstruct \textit{all} individual couplings, not just the mean value examined in prior work.

To better evaluate the statistical properties of estimated $\{J_{ij}\}$, we plot the histogram $\rho(J_{ij})$ for the three samples in Fig.~\ref{fig:compression_histogram}(a). For $\lambda/D = 1.11$, $\rho(J_{ij})$ yields a mean $\langle J_{ij}\rangle/k_BT \simeq -0.14$ and standard deviation $\sigma_J/k_BT \simeq 0.09$. Although most $J_{ij}$ values are negative, a small fraction of $J_{ij}$ in the right tail exhibit positive values.

For the smaller lattice spacing $\lambda/D = 1.10$, $\rho(J_{ij})$ spreads out significantly and its peak position shifts toward the left (more negative). This trend continues with the smallest lattice spacing $\lambda/D = 1.09$, where the peak of $\rho(J_{ij})$ moves further to the left and its width broadens further.

Figure~\ref{fig:compression_histogram}(b) plots $\langle J_{ij}\rangle$ and $\sigma_J$ for the three compressed samples. All $\langle J_{ij}\rangle$ values are negative, indicating antiferromagnetic couplings, as expected \cite{2008.han,2009.shokef,2023.hill}. However, the increase in $\sigma_J$ upon decreasing lattice spacing is a new property inaccessible to previous work. The increase in $\sigma_J$ suggests that as the packing fraction of the buckled particle system increases, the nearest-neighbor interactions, as characterized by $J_{ij}$, become more heterogeneous. Since $\{J_{ij}\}$ are fixed on the lattice, their distribution characterizes quenched disorder. We attribute the quenched disorder to particle size polydispersity and nonuniform local deformation observed in these buckled colloidal samples \cite{2008.han,2009.shokef,2023.hill}.

\subsection{Estimated couplings from sheared samples}
In the shear experiments, we have observed stripes of particles in the same buckled state as the shear strain $\epsilon$ increases. This can be qualitatively understood through the relief of geometric frustration by shear, as additional space along the $60^\circ$ direction is created, while along the $120^\circ$ direction particles are more closely packed. Consequently, particles tend to share the same buckled state along the $60^\circ$ direction. In terms of an effective Ising model, we thus expect spatially anisotropic spin couplings in the reconstructed parameters. To verify, we apply MLE to sheared samples with $\epsilon = 0.0\%$, $5.3\%$, and $7.9\%$. Again, we first confirm that $\sigma_J$ plateaus within the data duration for all datasets.

\begin{figure}[htbp]
	\centering
	\includegraphics[width=1.0\linewidth]{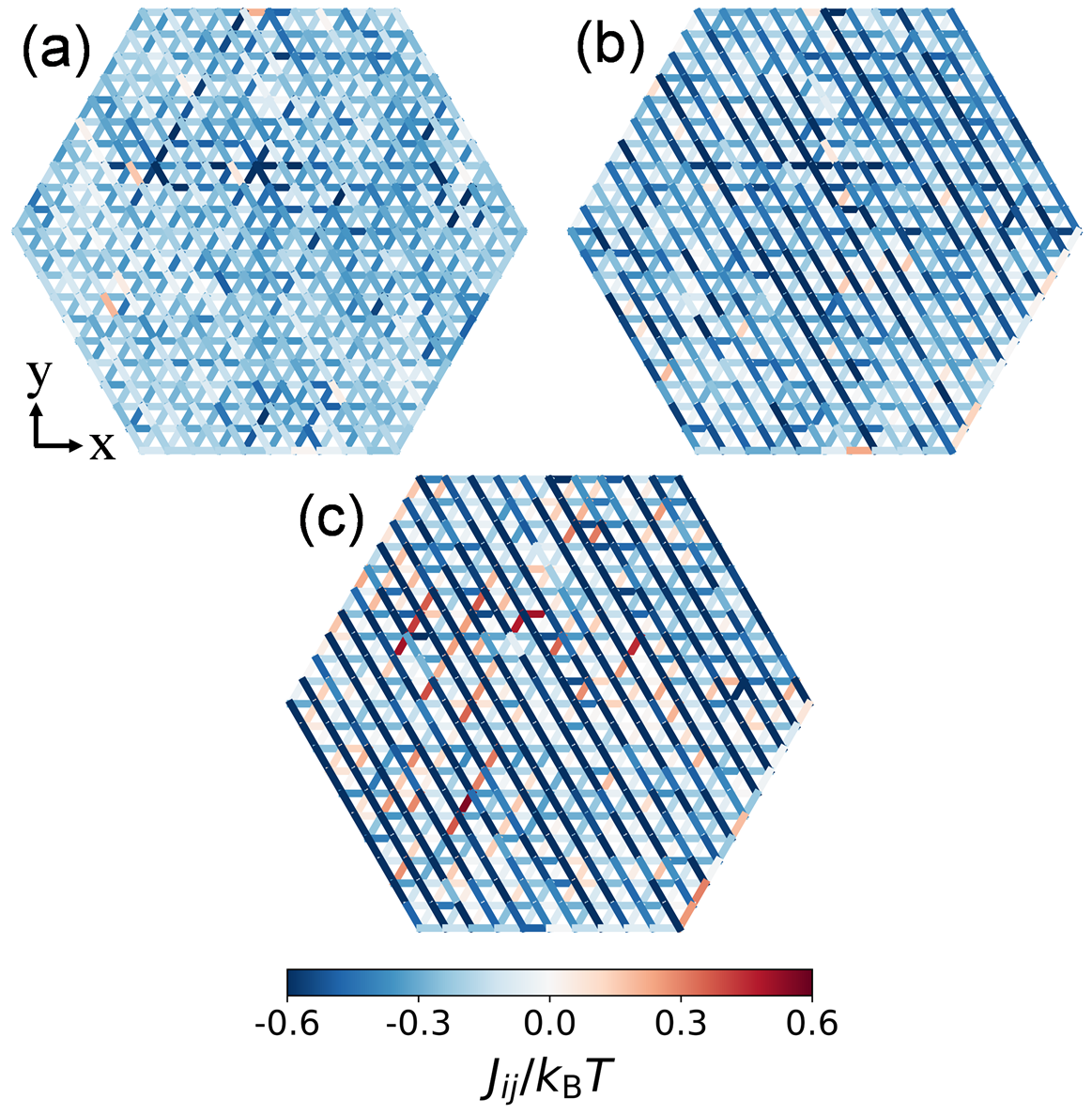}
	\caption{Spatial arrangement of estimated $\{J_{ij}\}$ from sheared samples for (a) $\epsilon = 0.0\%$, (b) $\epsilon = 5.3\%$, and (c) $\epsilon = 7.9\%$.}
	\label{fig:shear}
\end{figure}

Figure~\ref{fig:shear} plots the spatial arrangement of estimated $\{J_{ij}\}$ for the three shear samples. As $\epsilon$ increases from $0.0\%$ to $7.9\%$ [Figs.~\ref{fig:shear}(a) and (c)], the reconstructed couplings exhibit clear spatial anisotropy, with stripe-like patterns of same-colored bars gradually emerging. Along the $120^\circ$ direction, the inferred $\{J_{ij}\}$ become strongly negative (dark blue), while along the $60^\circ$ direction, the inferred $\{J_{ij}\}$ have smaller magnitudes with some gaining positive (red) values.

\begin{figure}[htbp]
	\centering
	\includegraphics[width=0.7\linewidth]{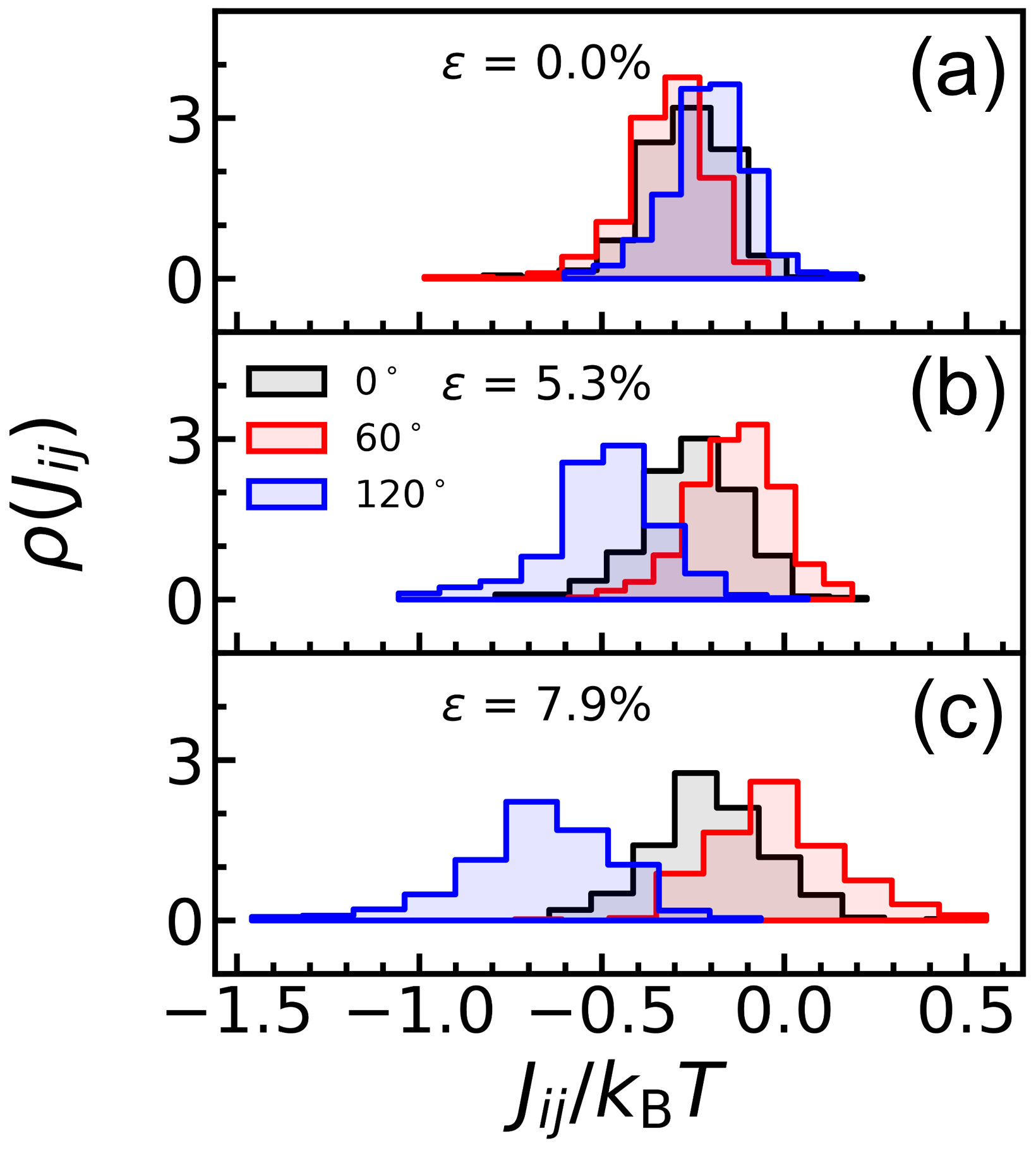}
	\caption{(a) Histograms of estimated couplings $\rho(J_{ij})$ for sheared samples, computed from bonds of the same direction ($0^\circ$ (black), $60^\circ$ (red), and $120^\circ$ (blue)). (b) Mean $\langle J_{ij}\rangle$ and standard deviation $\sigma_J$ of the orientation-dependent $\rho(J_{ij})$.}
	\label{fig:shear_histogram}
\end{figure}

To characterize the anisotropic spatial arrangement of couplings, we separate $\{J_{ij}\}$ along the three principal lattice directions in the histograms to reveal the direction-dependent distribution. As shown in Fig.~\ref{fig:shear_histogram}, $\rho(J_{ij})$ exhibits distinct trends along the three directions as $\epsilon$ increases. $\rho(J_{ij})$ shifts to more negative values along the $120^\circ$ direction, while becoming more positive along the $60^\circ$ direction. Along the $0^\circ$ direction, $\rho(J_{ij})$ exhibits a negligible shift as compared to the other two directions.

\begin{figure}[htbp]
	\centering
	\includegraphics[width=0.7\linewidth]{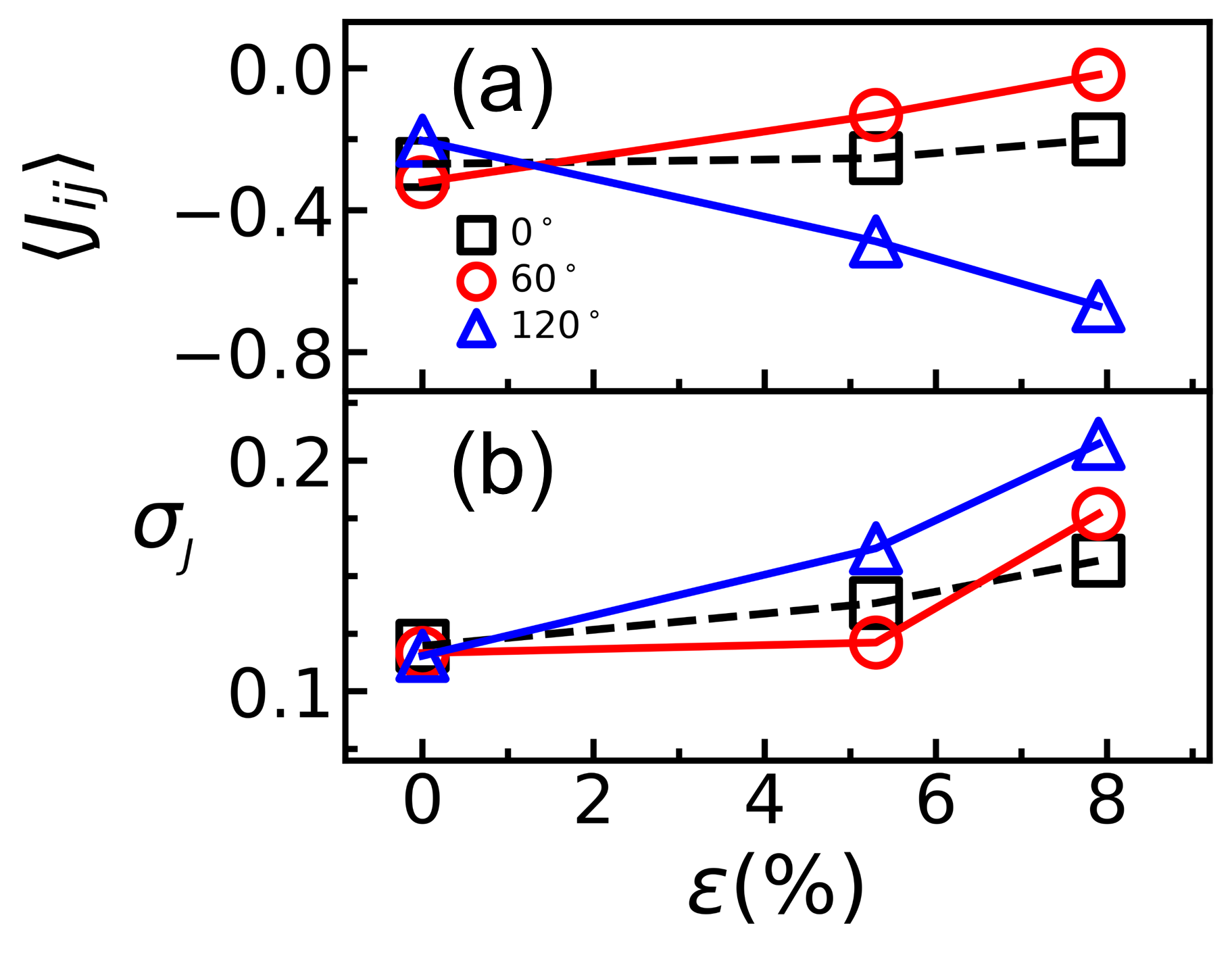}
	\caption{(a) Histograms of estimated couplings $\rho(J_{ij})$ for sheared samples, computed from bonds of the same direction ($0^\circ$ (black), $60^\circ$ (red), and $120^\circ$ (blue)). (b) Mean $\langle J_{ij}\rangle$ and standard deviation $\sigma_J$ of the orientation-dependent $\rho(J_{ij})$.}
	\label{fig:shear_mu_sigma}
\end{figure}

Figure~\ref{fig:shear_mu_sigma} plots $\langle J_{ij}\rangle$ and $\sigma_J$ for the direction-dependent $\rho(J_{ij})$. Along the $0^\circ$ direction, $\langle J_{ij}\rangle$ remains slightly negative and changes little with $\epsilon$. Along the $60^\circ$ direction, $\langle J_{ij}\rangle$ increases toward positive values with $\epsilon$, indicating that antiferromagnetic couplings weaken and eventually cross over to ferromagnetic couplings. By comparison, $\langle J_{ij}\rangle$ along the $120^\circ$ direction becomes more negative with increasing $\epsilon$, reflecting enhanced antiferromagnetic coupling. These inferred $\{J_{ij}\}$ suggest that one can use the shear direction to tune direction-dependent coupling parameters.

Unlike $\langle J_{ij}\rangle$, $\sigma_J$ along all three orientations increases monotonically with $\epsilon$. This uniform growth in $\sigma_J$ indicates that shear deformation induces stronger quenched disorder regardless of coupling bond direction.

\subsection{Estimated couplings from samples with tunable attractions}
Finally, we apply MLE to attraction-tuning experiments. As the attraction increases monotonically with sample temperature, we present results at $T = 24\,^{\circ}\mathrm{C}$, $26\,^{\circ}\mathrm{C}$, and $28\,^{\circ}\mathrm{C}$ to demonstrate the dependence of coupling parameters on attraction strength. For each temperature, we first verify that $\sigma_J$ plateaus within the dataset durations.

\begin{figure}[htbp]
	\centering
	\includegraphics[width=1.0\linewidth]{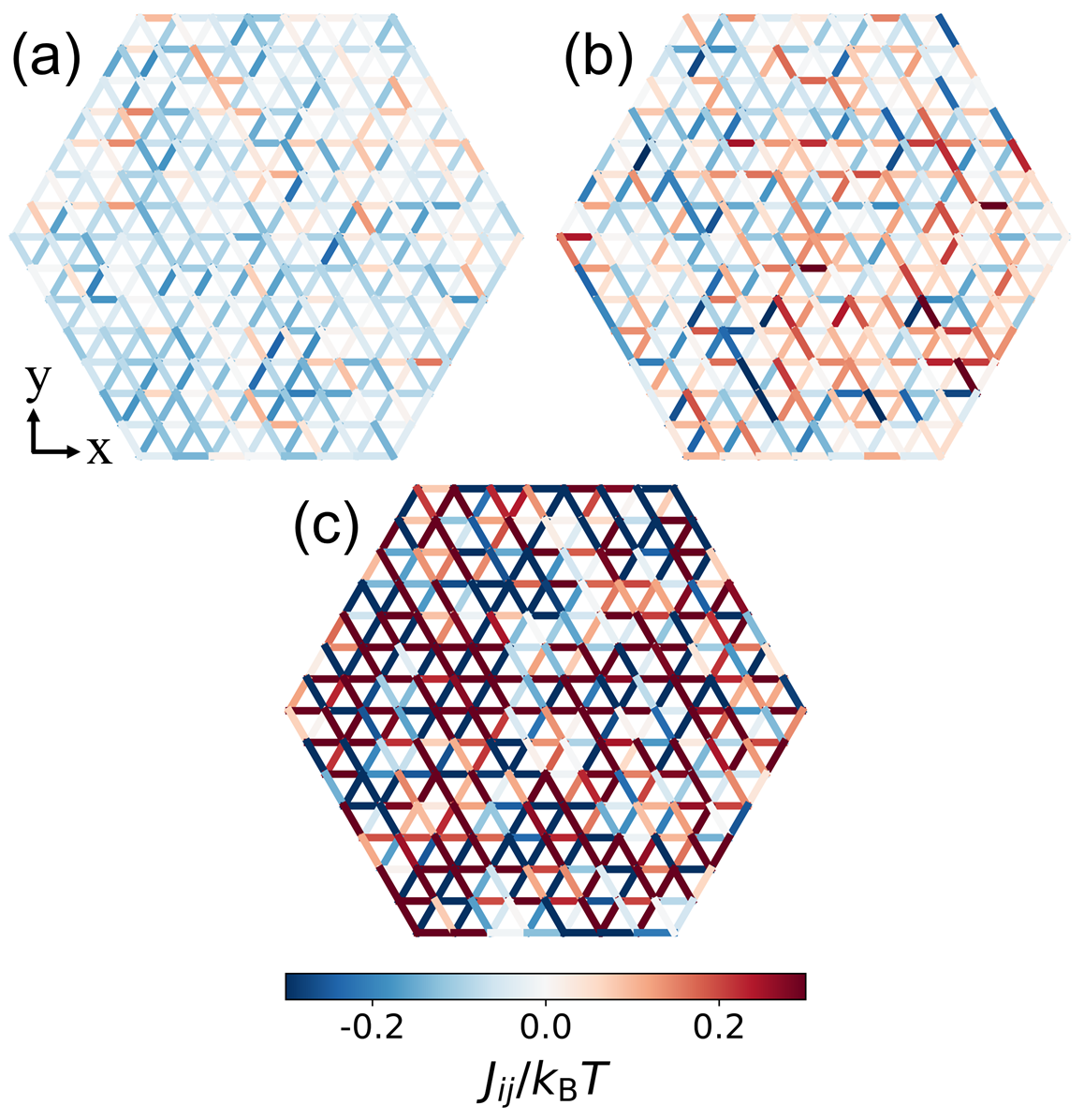}
	\caption{Spatial arrangement of estimated $\{J_{ij}\}$ from samples with tunable attractions at (a) $T = 24\,^{\circ}\mathrm{C}$, (b) $T = 26\,^{\circ}\mathrm{C}$, and (c) $T = 28\,^{\circ}\mathrm{C}$.}
	\label{fig:temperature}
\end{figure}

Figure~\ref{fig:temperature} plots the spatial arrangement of estimated $\{J_{ij}\}$ for samples at three temperatures. For $T = 24\,^{\circ}\mathrm{C}$, both positive and negative $J_{ij}$ values can be identified in the spatial map. At this temperature, the magnitudes of $\{J_{ij}\}$ are close to zero, shown by the light blue and light red colors [Fig.~\ref{fig:temperature}(a)]. From the map, negative couplings (blue) appear to dominate over positive ones (red) in terms of number.

At the higher temperature $T = 26\,^{\circ}\mathrm{C}$, the magnitudes of $\{J_{ij}\}$ of both signs (colors) exhibit a clear increase from $T = 24\,^{\circ}\mathrm{C}$ [Fig.~\ref{fig:temperature}(b)], suggesting the range of $\{J_{ij}\}$ values is broadened by the increase in attraction strength. This trend continues to the highest temperature $T = 28\,^{\circ}\mathrm{C}$, where $\{J_{ij}\}$ values are represented by much darker blue and red bars across the map. It also seems that the number of positive $J_{ij}$ surpasses that of negative ones, with the overall couplings dominated by the positive couplings. 

\begin{figure}[htbp]
	\centering
	\includegraphics[width=1.0\linewidth]{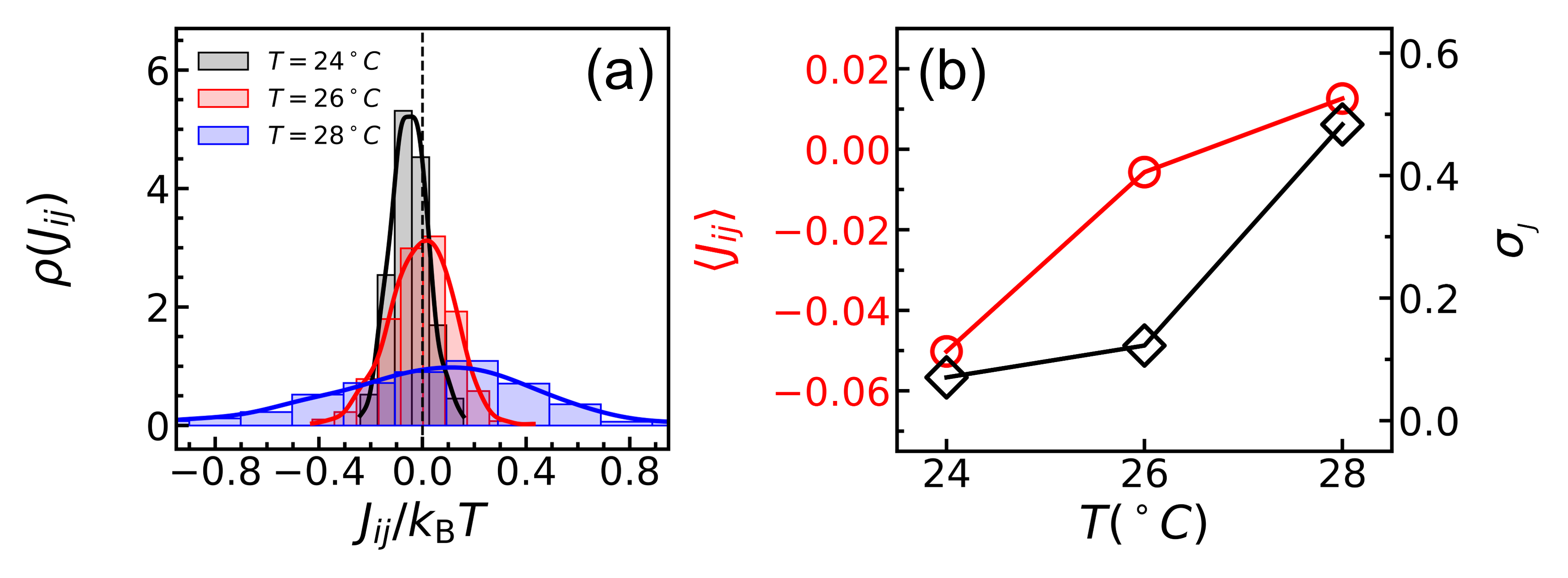}
	\caption{(a) Histograms of estimated couplings $\rho(J_{ij})$ for samples with tunable attractions. (b) Mean $\langle J_{ij}\rangle$ and standard deviation $\sigma_J$ of $\rho(J_{ij})$.}
	\label{fig:temperature_histogram}
\end{figure}

Figure~\ref{fig:temperature_histogram}(a) plots $\rho(J_{ij})$ for the three samples. For the lowest temperature (minimal attractive forces), $\rho(J_{ij})$ is sharply distributed around a slightly negative mean. As temperature, or equivalently attraction strength, rises, $\rho(J_{ij})$ significantly broadens and its peak clearly moves toward the positive direction. Figure~\ref{fig:temperature_histogram}(b) plots the mean $\langle J_{ij}\rangle$ and standard deviation $\sigma_J$ vs temperature. $\langle J_{ij}\rangle$ is found to increase monotonically from negative to positive with increasing temperature, showing a crossover from predominantly antiferromagnetic to ferromagnetic couplings. This result corroborates prior work where the mean coupling energy was derived from particle free energies \cite{2023.hill}. Meanwhile, $\sigma_J$ increases from $0.07\,k_BT$ at $24\,^{\circ}\mathrm{C}$ to $0.48\,k_BT$ at $28\,^{\circ}\mathrm{C}$, indicating stronger quenched disorder driven by increased attraction. The broadened distribution of $\{J_{ij}\}$ cannot be captured by the method used before \cite{2023.hill}.

\section{Summary}
\label{sec:summary}
In this work, we infer nearest-neighbor spin coupling distributions directly from experimental data. Specifically, we apply maximum likelihood estimation, a standard method for the inverse Ising problem, to a buckled colloidal monolayer system. Using synthetic simulation data with known ground truth, we establish that the standard deviation of the estimated couplings converges monotonically to its true value as the dataset size increases, and propose this convergence as a practical, ground-truth-free criterion for assessing inference reliability, which is an essential capability for real experiments where true couplings are inaccessible. We show that accurate estimation requires a sufficient number of equilibrium samples. As an example, in one of our experiments, the required recording time is about 100 times longer than the system relaxation time. This approach resolves individual nearest-neighbor spin couplings, providing a higher level of detail than previous studies.

With access to the spin coupling distribution, we characterize how experimental controls affect the system. Under isotropic compression, interaction tuning, and shear, we observe changes in heterogeneity through the coupling distributions rather than through averaged quantities. Isotropic compression systematically shifts the mean coupling to more negative (antiferromagnetic) values while broadening the distribution; temperature-dependent depletion attraction drives the mean coupling from negative to positive values, marking a crossover from repulsive to attractive effective interactions. In attraction-tuning experiments, we distinguish ferromagnetic from antiferromagnetic phases based on coupling types. In shear experiments, we examine spatial structures, such as zigzag stripes, using spatial maps, and resolve the anisotropic response of couplings along the three principal lattice directions, where shear relieves frustration along the $60^\circ$ orientation while compressing the $120^\circ$ orientation, producing direction-dependent coupling signs. Access to the coupling distribution therefore strengthens the role of buckled colloidal monolayers as a model system for studying disorder and phase behavior, and as a platform for computation based on the Ising model.

Although we demonstrate this method on a buckled colloidal monolayer, the framework can be extended to other artificial spin networks. Buckled colloidal monolayers benefit from thermal fluctuations and relatively fast equilibration, making them suitable for applying basic inverse Ising methods. Our benchmark analysis further indicates that reliable MLE reconstruction requires the characteristic coupling energy to remain comparable to or smaller than the thermal energy ($|J_{ij}|/k_BT \lesssim 1$); for systems with strongly ordered couplings ($|J_{ij}|/k_BT \gg 1$), the reduced configurational entropy and longer autocorrelation times degrade estimation accuracy, and more advanced inference schemes may be necessary. In contrast, for systems that do not readily reach equilibrium, more advanced inference methods beyond maximum likelihood estimation may be required.

\section*{Acknowledgement}
We thank Yanqing Hu for helpful discussions. We thank the National Key Research and Development Program of China (Grant No. 2022YFA1405002), National Natural Science Foundation of China (Grant No. 12274195), and Department of Science and Technology of Guangdong Province (Grant No. 2021QN02C382).

\section*{Data Availability}
The data that support this investigation are available from the authors upon reasonable request.
\bibliography{references}

\end{document}